%% file: main.tex
\documentclass[twocolumn]{revtex4}
\usepackage[utf8]{inputenc}
\usepackage[T1]{fontenc}

\usepackage{graphicx}
\usepackage{dcolumn}
\usepackage{bm}
\usepackage{amsmath}
\usepackage{amsthm}
\usepackage{hyperref}
\usepackage{braket}
\usepackage{tikz}
\usepackage{lipsum}
\usepackage{float}
\usepackage{blkarray}
\usepackage{amssymb}
\usepackage{dsfont}

\newtheorem{definition}{Definition}[section]
\newcommand{\matindex}[1]{\mbox{\scriptsize#1}}

\begin{document}

\title{A coherent approach to quantum-classical optimization}

\input{sec/00-authors}
\input{sec/01-abstract}

\maketitle

\input{sec/02-introduction}

\input{sec/03-methods}
\input{sec/04-results_discussions}

\input{sec/05-acknowledgements}
\input{sec/06-data_availability}

\input{sec/07-author_contributions}

\input{main.bbl}

\input{sec/08-appendix}

\end{document}

%% file: sec/00-authors.tex
\author{Authors: Andrés N. Cáliz 
\footnote{\texttt{andres.navas@qilimanjaro.tech}} $^1$ Jordi Riu \footnote{\texttt{jordi.riu@qilimanjaro.tech}} $^{1,2}$, Josep Bosch$^1$, Pau Torrente$^1$, Jose Miralles$^1$, Arnau Riera$^1$}

\affiliation{Qilimanjaro Quantum Tech, 08019 Barcelona, Spain.$^1$}
\affiliation{Departament de Física, Universitat Politècnica de Catalunya,
Barcelona 08034, Spain.$^2$}
\date{\today}

%% file: sec/01-abstract.tex
\begin{abstract}
Hybrid quantum-classical optimization techniques, which incorporate the pre-optimization of Variational Quantum Algorithms (VQAs) using Tensor Networks (TNs), have been shown to allow for the reduction of quantum computational resources. In the particular case of large optimization problems, commonly found in real-world use cases, this strategy is almost mandatory to reduce the otherwise unfathomable execution costs and improve the quality of the results. We identify the coherence entropy as a crucial metric in determining the suitability of quantum states as effective initialization candidates. Our findings are validated through extensive numerical tests for the Quantum Approximate Optimization Algorithm (QAOA), in which we find that the optimal initialization states are pure Gibbs states. Further, these results are explained with the inclusion of a simple and yet novel notion of expressivity adapted to classical optimization problems. Based on this finding, we propose a quantum-classical optimization protocol that significantly improves on previous approaches for such tasks, with specific focus on its effectiveness.
\end{abstract}

%% file: sec/02-introduction.tex
\section{Introduction}
\label{sec:introduction}

In an increasingly interconnected world, where individuals, companies, and international organizations must optimize resources to achieve maximum results, the ability to effectively solve optimization problems is becoming a critical skill. By optimizing processes, minimizing costs, and maximizing profits, organizations can not only improve efficiency but also significantly impact profitability and sustainability. This capability offers a strategic advantage in a competitive landscape where efficiency and agility are essential. In the field of quantum computing, the potential to revolutionize optimization techniques is especially exciting. Quantum algorithms, which take advantage of the unique properties of quantum mechanics, are expected to be able to solve complex optimization problems more efficiently than classical algorithms. This ability to outperform traditional methods positions quantum computing as a key player in the future of optimization.\\

However, despite fast progress in the coming years quantum computers are expected to remain in an early stage of development, characterized by a limited number of low-quality qubits, the high cost and scarcity of quantum resources compared to classical computing resources. This situation has prompted the development of variational quantum algorithms, which are specifically designed to leverage the currently available, albeit noisy, quantum resources \cite{cerezo_2021, bharti_2022, endo_2021}. Within this framework, quantum devices are being explored for tackling complex problems in quantum chemistry \cite{cao_2019, mcArdle_2020, lubasch_2020}, quantum machine learning \cite{biamonte_2017, perdomo_Ortiz_2018, huang_2022, havl_ek_2019}, and other areas where classical algorithms face significant challenges. Despite their potential, there remains considerable debate regarding the actual advantages that these hybrid algorithms can provide \cite{anschuetz_2022}. Key challenges facing VQA's include the issue of \textit{barren plateaus} \cite{cerezo_2024, marrero_2021, wang_2021, holmes_2021}, where the optimization landscape becomes exponentially flat, impedes effective learning, and gives the algorithms the tendency to become trapped in local minima \cite{arrasmith_2022, anschuetz_2023}. \\

To enhance the performance of variational quantum algorithms, several strategies have been developed to provide more advantageous starting points for locating higher quality minima. One such strategy involves the intelligent initialization of the parameters associated with the parameterized quantum circuits (PQC) used in VQA's, which helps mitigate the occurrence of barren plateaus, thereby improving the algorithm's ability to find optimal solutions \cite{zhang_2022, park_2024, wang_2023, park_2024_hea, shi_2024}. Other strategies use classical methods such as semidefinite programming to generate solutions that serve as initial points for VQAs \cite{egger_2021}. Additionally, iterative VQA approaches that start in energetically favorable regions further enhance the efficiency and effectiveness of the algorithm \cite{puig_2024}. There are also approaches that focus their analysis on the cost function to be optimized to improve VQA performance \cite{li_2020}. Finally, another promising approach is to initialize VQA's with energetically favorable initial states, exploiting quantum phenomena targetted at the hard part of the optimization whilst reducing its computational burden. Several Tensor Networks techniques have been employed to construct these beneficial initial states \cite{huggins_2019, dborin_2021, rudolph_2023}. Among these techniques, the optimization scheme proposed in Ref. \cite{rudolph_2023} is one of the most advanced and complete by combining tensor network methods with quantum optimization algorithms. \\

\input{figure/itevo_qaoa_protocol_final}

However, this scheme is not functional when applied to classical problems because the TN algorithm produces initial states close to those of the computational basis. Then the VQA algorithm is not efficient in generating best energy solution states (for more details, see Appendix \ref{app:synergetic_training}). \\

In this work, we implement a new classical-quantum optimization scheme in which we first identify the most suitable quantum states to initialize quantum variational algorithms, which are found to be pure Gibbs states. From this, we adapt the Tensor Network techniques in order to generate them for industrial optimization problems. We specifically focus on the Quantum Approximate Optimization Algorithm \cite{farhi_2014}, as it is one of the least intensive ansatz in terms of parameters, while allowing for a theoretical understanding behind the numerical advantage. Nevertheless, the numerical results suggest that the approach seems to work quite well in general for different types of ansatz. \\

A detailed depiction of our proposed scheme is shown in \mbox{Fig. \ref{fig:protocol_itevo_qaoa}}. First, as shown in Figure \mbox{Fig. \ref{fig:protocol_itevo_qaoa}} (a) the process starts with the use of a TN algorithm called $\text{MPO} \; W^{\text{I,II}}$ \citep{zaletel_2015} to approximately construct the desired pure Gibbs state, \mbox{encoded} via a Matrix Product State (MPS). Secondly, and as can be seen in Figure \mbox{Fig. \ref{fig:protocol_itevo_qaoa}} (b), a decoding algorithm \cite{rudolph_2022} is applied to transform the MPS state into a state preparation circuit through a set of gates \textit{SU(4)}. These gates \textit{SU(4)} can be physically implemented using parameterized quantum gates designed for 1- and 2-qubit systems, employing the decomposition protocol \textit{KAK} \cite{tucci_2005}. The state preparation circuit, which approximately generates the Gibbs states, will remain frozen during VQA execution. This is done to minimize the number of parameters to be optimized.  Finally, as shown in Figure \mbox{Fig. \ref{fig:protocol_itevo_qaoa}} (c) we add a series of QAOA layers to the previously generated ansatz circuit, whose parameters are then optimized using evolutionary or gradient-based techniques, thus completing the optimization protocol.

%% file: figure/itevo_qaoa_protocol_final.tex
\begin{figure*}[!t]
\centering
\includegraphics[width=2\columnwidth]{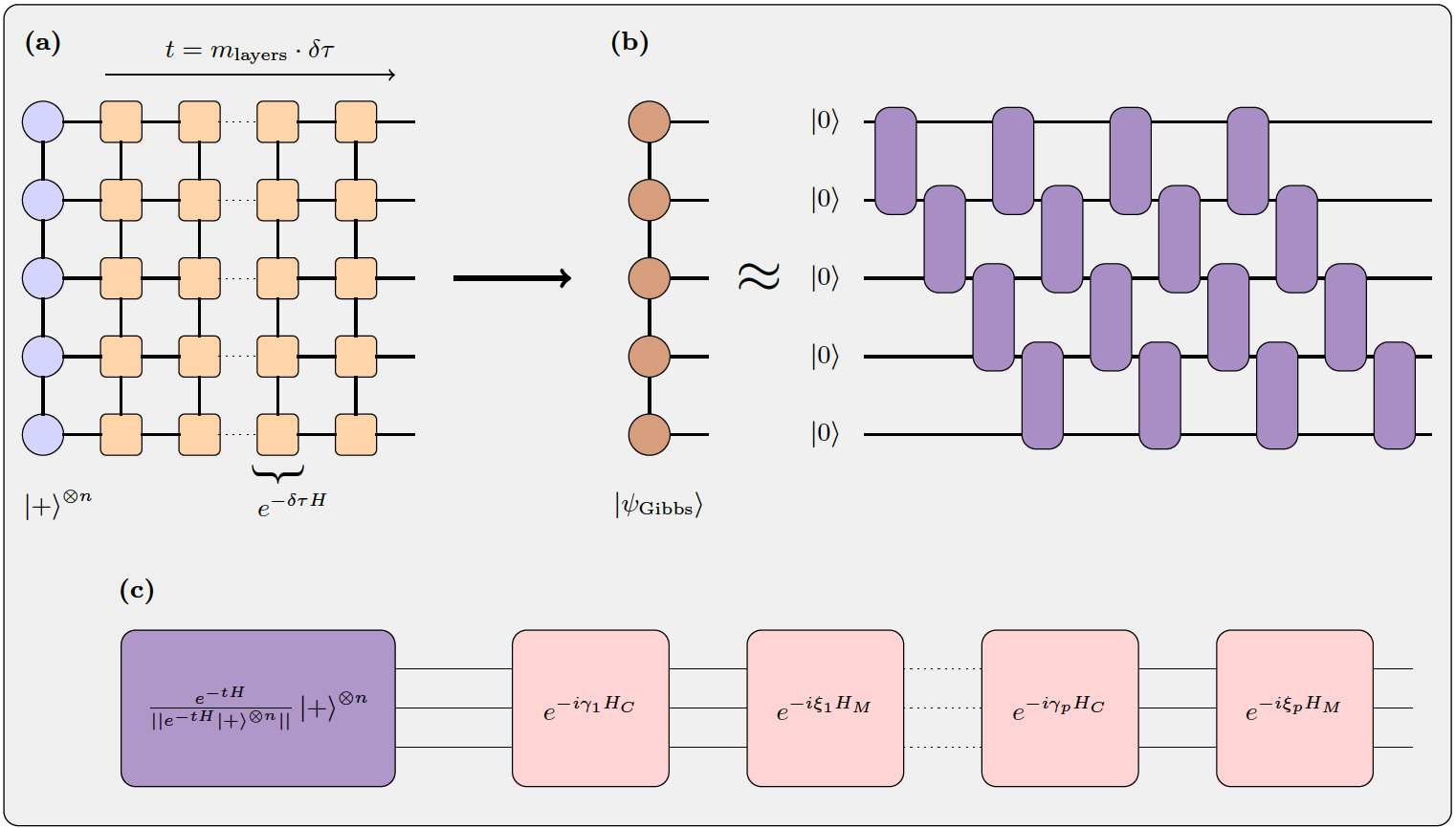} 
\caption{Schematic representation of the combined training framework using imaginary time evolution in tensor networks and the Approximate Quantum Optimization Algorithm. (a) The pure Gibbs state is generated through imaginary time evolution. Each layer of the Matrix Product Operator (MPO) advances the initial Hadamard state by an imaginary time increment, $\delta \tau$, resulting in the final pure Gibbs state at a temperature $T = \frac{1}{m \cdot \delta \tau} = \frac{1}{t}$ where $t$ is the total evolution time and $m$ is the number of MPO layers applied. (b) The Gibbs state is then decomposed into a set of quantum gates \textit{SU(4)}, which can be further decomposed into parameterized quantum gates involving 1- and 2-qubit operations. (c) The quantum circuit that reproduces the Gibbs state is subsequently used for state preparation, serving as the initialization for the QAOA algorithm.}    
\label{fig:protocol_itevo_qaoa}  
\end{figure*}

%% file: sec/03-methods.tex
\section{Methods}
\label{sec:methods} 

\subsection{Diagonal Entropy as a Measure of Coherence}  
\label{subsec:coherence-measure}

In this section, we introduce the relative entropy of coherence as a measure of the coherence (or superposition) of a quantum state with respect to a given basis. \\

For any quantum state $\rho$, the relative entropy of coherence is defined as the shortest distance (measured by relative entropy) between the state \(\rho\) and its closest incoherent state. Mathematically, this is given by:

\begin{equation}\label{eq:coherence-entropy-a}
    S_C(\rho)= \min_{\sigma\in I}S(\rho||\sigma),
\end{equation}

\setlength{\parindent}{0pt}
where the minimization is performed over the set of incoherent states \(I\), and 
\(S(\rho||\sigma)=\text{Tr}\left(\rho \log_2 \rho - \rho \log_2 \sigma\right)\) represents the relative entropy between \(\rho\) and \(\sigma\). Given a reference basis \(\left\{ \ket{s}\right\}_{s=1}^{d}\), which in our case will be the computational basis, it can be shown that the incoherent state minimizing the distance from \(\rho\) in Eq.~\eqref{eq:coherence-entropy-a} is the completely dephased state of $\rho$ in the \(\left\{ \ket{s}\right\}_{s=1}^{d}\) basis,

\begin{equation}
    \rho_\textrm{d} = \sum_s \bra{s}\rho \ket{s} \ket{s}\bra{s}\, .
\end{equation}

Using this fact, and applying the definition of relative entropy, the relative entropy of coherence can be rewritten as:

\begin{equation}
S_{C}(\rho) = S(\rho_\textrm{d}) - S(\rho),
\label{eq:relative_coherence_entropy}
\end{equation}

\setlength{\parindent}{0pt}
where \(S(\rho)=-\text{Tr}\left(\rho\log_2\rho\right)\) is the Von Neumann entropy. For the purposes of this work, where we deal exclusively with pure states, the coherence entropy simplifies to the diagonal entropy, expressed as:

\begin{equation}
S_{C}(\rho) = S(\rho_\textrm{d}) =: S_d(\rho)\, .
\label{eq:relative_coherence_entropy_pure}
\end{equation}

\subsection{Pure Gibbs State as the Maximizer of Relative Coherence Entropy (Given Energy)}
\label{subsec:gibbs_states}

One of the claims of the present paper is that coherence in the initial state of the VQA is beneficial for its performance. With this in mind, it is interesting that the cost function used in the pre-training tensor network method, which generates the initial state of the VQA, not only considers energy but also accounts for the amount of coherence in the state, measured by the diagonal entropy. This leads to a cost function for the pre-training method defined as
\begin{equation}
    G(\ket{\psi}) = \bra{\psi}H\ket{\psi} - T S_{d}(\ket{\psi}),
    \label{eq:gibbs_energy_free}
\end{equation}
where $T$ is a coefficient that controls the weight that has been given to the entropy in the cost function.
Note that if we interpret the diagonal entropy $S_d$ as the thermodynamic entropy in the above's expression \eqref{eq:gibbs_energy_free}, the cost function $G$ becomes the Helmholtz free energy and $T$ the temperature of the available bath. \\

It is known from statistical physics that the states that minimize the equation \ref{eq:gibbs_energy_free} are those that follow the Boltzmann distribution, as defined by 
\begin{equation}
    \ket{\psi(t)}=\frac{1}{\sqrt{Z_t}}\sum_s e^{-t E_s/2} e^{i\theta_s}\ket{s},
    \label{eq:gibbs_states}
\end{equation}
where $t=1/T$ is the inverse of the temperature, 
$Z_{t}=\sum_{s} e^{-t E_s}$ is the partition function of the Hamiltonian and $\theta_s$ is a set of arbitrary phases which for simplicity we will assign the value of $\theta_s=0$ for all bit strings $s$.

\subsection{Energy - Entropy Diagram}
It is useful to introduce the energy-entropy diagram, as employed in thermodynamics papers like the one presented in Ref. \cite{bera_2019}. For a system described by a Hamiltonian $H$, a state $\ket{\psi}$ is represented on the energy-entropy diagram by a point with coordinates  \mbox{$x(\psi) = (E(\psi), S_{d}(\psi))$}, as illustrated in Figure \ref{fig:energy_entropy_diagram}. This entropy-energy diagram uses the diagonal entropy and is restricted to pure states. In this diagram, all these states are confined in the region bounded by the horizontal axis, where $S_d=0$ corresponding to the eigenstates of the Hamiltonian, and a convex curve, $(E(t),S(t))$, representing the pure Gibbs states with maximum entropy at both positive and negative temperatures. We refer to this curve as the \textit{Boltzmann limit}. The inverse temperature associated to one point of the Boltzmann limit is given by the slope of the tangent line in such a point, which is given by 

\begin{equation}
  t = \frac{d S(t)}{d E(t)}.
  \label{eq:partial_beta}
\end{equation}

\input{figure/energy_entropy_diagram}

\subsection{MPO $W^{\text{I,II}}$ Algorithm}

There are different approaches to generate pure Gibbs states using tensor networks, the most prominent being the time-evolving block decimation (TEBD) \cite{paeckel_2019}. However, the Hamiltonians of our interest typically contain non-local interactions since they are classical Hamiltonians and so we have chosen to employ the methods described in Ref. \citep{zaletel_2015}, where the evolving blocks are represented by Matrix Product Operators (MPOs) that approximate the time-independent Hamiltonian evolution operator $e^{- i \delta t H}$. By denoting $i \delta t$ as $\delta \tau$, the operator for $\text{MPO} \; W^{\text{I,II}}$ becomes $e^{- \delta \tau H}$, representing the imaginary time evolution of the quantum system. In general, the problem of encoding Hamiltonians containing non-local interactions in an MPO  is a non-trivial task. To encode Hamiltonians of classical Ising-type problems with long-range interactions in an MPO form, the methods presented in Ref. \cite{frwis_2010} can be used. Consider a Hamiltonian that can be decomposed into a sum of local terms of the form $H = \sum_{x} H_{x}$, which can be rewritten as 

\begin{equation}
  H = H_{L_i} \otimes \mathds{1}_{R_i} + \mathds{1}_{L_i} \otimes H_{R_i} + \sum_{a_i=1}^{N_i} h_{L_i, a_i} \otimes h_{R_i, a_i}.
  \label{eq:hamiltonian}
\end{equation}

\setlength{\parindent}{0pt}
Here, $H_{L_i}$ and $H_{R_i}$ represent the components of the Hamiltonian located to the left and right of the link at position $i$. The terms $h_{L_i, a_i} \otimes h_{R_i, a_i}$ correspond to the $N_i$ interaction terms that cross through the link between two consecutive sites. An equivalent representation of equation \ref{eq:hamiltonian} is given by the expression

\begin{equation}
\begin{pmatrix}
H_{R_{i-1}} \\
h_{R_{i-1}, a_{i-1}} \\
\mathds{1}_{R_{i-1}}
\end{pmatrix}
=
\begin{blockarray}{rccc}
    & \matindex{1} & \matindex{$N_{i}$} & \matindex{1} \\
    \begin{block}{r(ccc)}
      \matindex{1 \;} & \hat{\mathds{1}} & \hat{C}_i & \hat{D}_i \\
      \matindex{$N_{i-1}$ \;} &  0 & \hat{A}_i & \hat{B}_i \\
      \matindex{1 \;} & 0 & 0 & \hat{\mathds{1}} \\
    \end{block}
\end{blockarray}
\;
\;
\otimes
\begin{pmatrix}
H_{R_i} \\
h_{R_i, a_i} \\
\mathds{1}_{R_i}
\end{pmatrix}
.
\label{eq:mpo_hamiltonian}
\end{equation}

The central tensor in the equation represents the MPO operator for the $i$-th position, corresponding to the Hamiltonian described in \ref{eq:hamiltonian}. The dimensionality of blocks $A_i, B_i, C_i, D_i$ acting on the $i$-th site is indicated on the expression. Given the $i$-th tensor of the MPO, the algorithm $\text{MPO} \; W^{\text{I, II}}$ implements an approximation of the time evolution operator expressed as

\begin{equation}
U(\delta\tau) = 1 + \delta\tau \sum_{x} H_x + \frac{1}{2} \delta\tau^2 \sum_{x,y} H_x H_y + \ldots
\label{eq:taylor_evolution_operator}
\end{equation}

$\text{MPO} \; W^{\text{I}}$ implements the approximation of the time evolution operator without considering any overlapping terms, while $W^{\text{II}}$ includes also terms that overlap in at most one site. Both approximations have an effective error of $\mathcal{O}(L\delta\tau^2)$, with $L$ being the system size, even though in most cases $W^{\text{II}}$ will give a more accurate representation as discussed in the original paper.
An $\text{MPO} \; W^{\text{I,II}}$ layer will bring an MPS representing $\ket{\psi_0}$ into a new quantum state $\ket{\psi_{\delta \tau}} = U(\delta\tau)\ket{\psi_0}$ (for more details on the implementation of the algorithm, see Appendix \ref{app:mpo_time_evolution})

\subsection{Pure Gibbs states from $\text{MPO} \; W^{\text{I,II}}$}

$\text{MPO} \; W^{\text{I,II}}$ can be used to construct pure Gibbs states. This is achieved through imaginary time evolution, where the evolution is governed by the operator $U(\delta \tau) = e^{- \delta\tau H}$ applied to the initial state \mbox{$\ket{\psi_0}=\ket{+}^{\otimes n}$} as shown in the following expression
\begin{equation}
\ket{\psi_{\delta \tau}}= \phi_{\delta t}\left(\ket{\psi_0}\right)=\frac{e^{-\delta \tau H}\ket{\psi_{0}}}{|| e^{-\delta \tau H}\ket{\psi_{0}}||}.   
\label{eq:gibbs_states_tevo}
\end{equation}
where $\phi_{\delta t}(\cdot )$ is a transformation that consists of the application of $U(\delta t)$ followed by a normalization. The MPO generated by the algorithm evolves the state by a single $\delta \tau$ per MPO layer. To perform a complete imaginary time evolution over a total time $t$, it is necessary to apply $m$ layers of $U(\delta \tau)$, or more precisely $m$ $\phi_{\delta t}(\cdot)$ transformations, which collectively evolve the system over a total time $t = m \cdot \delta \tau$. Note that since we are doing an imaginary time evolution, $U$ is no longer unitary and a normalization has to be performed after applying each MPO layer. The relationship between the temperature $T$ of the generated pure Gibbs state and the total imaginary evolution time $t$ is given by

\begin{equation}
t = \frac{1}{T}.  
\label{eq:time_temperature}
\end{equation}

In the limit of infinite $t$ (or $T=0$) , the ground state solution of the Hamiltonian is obtained, i.e.,

\begin{equation}
\lim_{t\to \infty}\ket{\psi_t}=\ket{E_0}.
\label{eq:ground_state}
\end{equation}

The trajectory followed by the system in the energy-entropy diagram lies precisely along the Boltzmann boundary. Its end-point depends on the maximum imaginary time evolution that can be simulated for the quantum state $\ket{\psi_{0}}$, which is determined by the maximum internal bond dimension $\chi_{max}$ allowed in the tensor network and the size of the time step $\delta \tau$ used (for more information, see Appendix \ref{app:mpo_gibbs_states}).

\subsection{MPS to PQC translation}

The conversion from a TN to a Parametrized Quantum Circuit is a well-known problem that can be effectively handled through classical simulation and the use of canonical forms \cite{perezgarcia_2007, shi_2006}. However, this direct translation from TN to PQC often results in the application of variable-range multi-qubit gates. These multi-qubit gates cannot be directly implemented on real hardware, and instead one needs to decompose them into a set of 1- and 2-qubit base gates, depending on the nature of the quantum hardware used \cite{lin_2022}. The decomposition, or transpilation, of multi-qubit gates into base gates is a non-trivial challenge itself. We employ the MPS to PQC decomposition presented in Ref. \cite{rudolph_2022}, which enables an efficient translation. This method combines analytical decomposition strategies \cite{ran_2020} with variational decomposition strategies \cite{shirakawa_2021}. Using this protocol, an MPS can be approximately converted into a PQC using $k$ layers of two-qubit $SU(4)$ quantum gates arranged in a staircase pattern. The value of $k$ is determined by the maximum allowable dimensionality $\chi_{max}$ of the MPS and the amount of entanglement present in the state contained within the MPS. Each general \( SU(4) \) gate is then decomposed into single-qubit and two-qubit gates that can be directly executed on quantum hardware, using the \textit{KAK} decomposition as described in Ref. \cite{tucci_2005}.

\subsection{Quantum Approximate Optimization Algorithm}

The Quantum Approximate Optimization Algorithm \cite{farhi_2014} is a variational quantum algorithm that is used to find the low-energy states of a problem Hamiltonian. The QAOA is structured in a sequence of layers, each of which includes two operators. The first operator, called the cost operator, is responsible for generating phase shifts between eigenstates according to their energy, while the second operator, called the mixer operator, is responsible for generating interferences by taking advantage of the phase shifts generated by the cost operator. In our case, the cost Hamiltonian will be given by

\begin{equation}
\label{eq:cost_hamiltonian}
    H_{C} = \sum_{i} h_i \sigma_{z}^{i} + \sum_{i, j>i} J_{ij} \sigma_{z}^{i} \sigma_{z}^{j},
\end{equation}

where $\sigma_{z}^{i}$ is the Pauli-z operator acting on qubit $i$. On the other hand, $h_i$ and $J_{ij}$ are the coefficients associated with the specific classical problem. The constant terms that typically arise when transforming the Quadratic Unconstrained Binary Optimization, QUBO, formulation into the problem's Hamiltonian, as shown in expression \ref{eq:cost_hamiltonian}, are generally disregarded. This is because they do not affect the final result, serving only as a shift in the reference of the energy levels. The parameterized quantum circuit of the QAOA algorithm is given by 

\begin{equation}
\label{eq:pqc_qaoa}
        U(\vec \gamma,\vec \xi)=\prod_{k=1}^p e^{-i \gamma_k H_C} e^{- i \xi_k H_M},
\end{equation}

$H_M$ is the so-called mixing Hamiltonian, and in its simplest version takes the form $H_M=-\sum_{i=1}^n \sigma_{x}^{i}$.  The product implies the application of $p$ layers of the operators $e^{-i \gamma_k H_C}$ and $e^{- i \xi_k H_M}$, which act on the initial state, typically $\ket{+}^{\otimes n}$. The set of parameters, denoted as $\vec{\gamma}$ and $\vec{\xi}$, is typically obtained using a classical optimizer (in this work, we have employed both the \textit{COBYLA} optimizer \cite{powell_1994} and the \textit{CMA-ES} optimizer \cite{hansen_2023}). These optimizers iteratively update the parameters in order to minimize the cost function, which is given by the expression

\begin{equation}
\label{eq:expected_value_qaoa}
        \min_{\vec{\gamma}, \vec{\xi}} \; \langle \psi(\vec{\gamma}, \vec{\xi}) | H_C | \psi(\vec{\gamma}, \vec{\xi}) \rangle,
\end{equation}

where $\langle \psi(\vec{\gamma}, \vec{\xi}) | H_C | \psi(\vec{\gamma}, \vec{\xi}) \rangle$ represents the expected value of the energy associated with the cost Hamiltonian. On the other hand, the state $\ket{\psi(\vec{\gamma}, \vec{\xi})}$ is the result of applying the operator given by the expression \ref{eq:pqc_qaoa} on an initial state as it appears in the following equation

\begin{equation}
\label{eq:state_qaoa}
        |\psi(\vec{\gamma}, \vec{\xi}) \rangle = U(\vec \gamma,\vec \xi) \ket{+}^{\otimes n}.
\end{equation}

%% file: figure/energy_entropy_diagram.tex
\begin{figure}[H]
\centering
\includegraphics[width=\columnwidth]{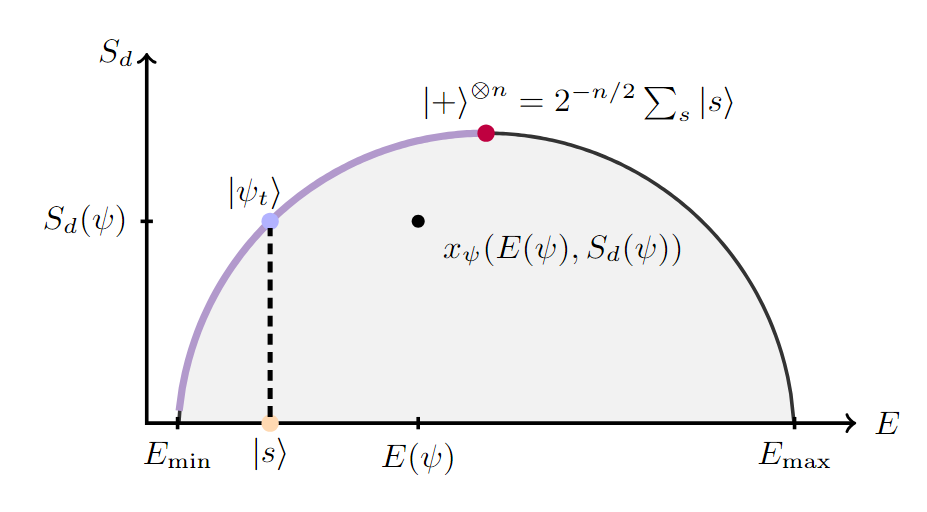} 
\caption{Energy-entropy diagram. Any quantum state $\psi$ is represented in the diagram as a point with coordinates $x_\psi:=(E(\psi), S(\psi))$. The states that minimize the Free Energy lie on the Boltzmann boundary (thick purple line). The ground state of the Hamiltonian, that encodes the solution of the problem, is in the lower-left corner of the Boltzmann boundary. The red dot represents the state with the highest entropy, known as the Hadamard state. States along the black line, including the orange and blue dots, represent states with the same energy but different entropy.}
\label{fig:energy_entropy_diagram}
\end{figure}

%% file: sec/04-results_discussions.tex
\section{Results and Discussions}
\label{sec:results_discussions}

\subsection{Study of the Impact of Coherence on the QAOA}
\label{subsec:results_gibbs_states}

We present a series of simulations aimed at comparing the effectiveness of pure Gibbs states, denoted as $\ket{\psi_{\text{Gibbs}}}$, against other candidate initial states. These states are categorized using the energy-entropy diagram shown in Figure \ref{fig:energy_entropy_diagram}. Specifically, we compare the performance of $\ket{\psi_{\text{Gibbs}}}$ with product states in the computational basis $\ket{s}$ of equal energy (Figure \ref{fig:energy_entropy_diagram}, orange dot), the Hadamard state $\ket{+}$ (Figure \ref{fig:energy_entropy_diagram}, red dot), and states with variable entropy states with energy values close or equal to those of the pure Gibbs states, denoted as Gaussian states $\ket{\psi_{\text{Gauss}}}$ (Figure \ref{fig:energy_entropy_diagram}, black dashed line). These Gaussian states are defined by the following expression

\begin{equation}
    \ket{\psi_{\text{Gauss}}(E_{T})} = \frac{1}{\sigma \sqrt{2 \pi}} \sum_{s}e^{-\frac{(E_{s} - E_{T})^2}{2 \sigma^2}} \ket{s},
    \label{eq:gaussian_states}
\end{equation}

where $E_T$ represents the expected energy of the quantum state and $\sigma$ the standard deviation of the distribution that impacts the effective number of quantum states participating in the linear combination. \\

All numerical tests are performed for random instances of the Max Cut \cite{lalovic_2024} problem. Given an undirected graph $ G = (V, E) $ with a set of vertices $ V $ and a set of edges $ E $, the objective of the Max Cut problem is to find a partition of the vertices $ V $ into two disjoint subsets $ A $ and $ B $ such that the number of crossing edges is maximum. The problem Hamiltonian associated with a given graph is given by the expression

\begin{equation}
    \text{minimize} \quad H =  - \frac{1}{2}\sum_{(i,j) \in E} (1 - \sigma_{z}^{i} \sigma_{z}^{j}).
    \label{eq:max_cut_hamiltonian}
\end{equation}

\begin{figure*}[t]
\centering 
\includegraphics[width=2.1\columnwidth]{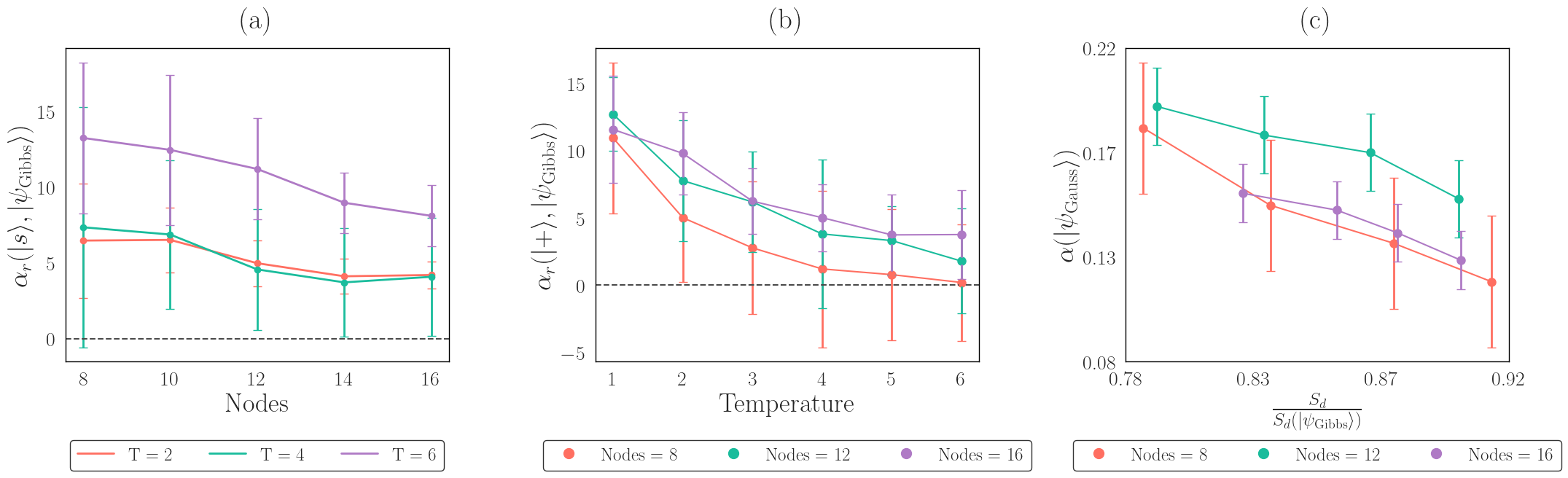} 
\caption{Approximation ratio results for different initialization states for the QAOA (a) Relative approximation ratio between pure Gibbs states and computational basis states of equal energy. (b) Relative approximation ratio between pure Gibbs states and the Hadamard state. (c) Approximation ratio for Gaussian states of equal energy and variable entropy, normalized with respect to the pure Gibbs state of the same energy.}
\label{fig:states_vs_gibbs_states} 
\end{figure*}

We initialize the parameters of the QAOA with values that are close to zero. This has been shown to help mitigate the effect of barren plateaus \cite{grant_2019}, thus while improving the convergence of the algorithm (although our results appear to be independent of the initialization strategy used for QAOA-associated angles).  \\

We evaluate the relative performance quality of the QAOA algorithm for two different initialization states $\ket{x}$ and $\ket{y}$ using the relative approximation ratio, which is defined as

\begin{equation}
\alpha_{r}(\ket{x}, \ket{y}) = \alpha(\ket{x}) - \alpha(\ket{y}) = \frac{E_{y} - E_{x}}{|E_{\textit{min}}|},
\label{eq:approx_ratio}
\end{equation}
where $\alpha(\ket{x})= (E_{\textit{min}} - E_{x})/|E_{\textit{min}}|$ represents the approximation ratio of a quantum state i.e. the quality of the solution relative to the ground state. For example, a positive value of $\alpha_{r}$ indicates that the quality of the $\ket{y}$ solution is higher than that of $\ket{x}$.\\

In Figure \ref{fig:states_vs_gibbs_states}, we present the results obtained from comparing the performance of pure Gibbs states with candidate states used to compete against them. Firstly, we observe that, even though it is positive, the relative difference in solution quality between initializing the QAOA with pure Gibbs states at different temperatures and product states of the same energy becomes progressively smaller as the problem size increases (Figure \ref{fig:states_vs_gibbs_states} (a)). This is due to the use of the relative quality metric of the solution with respect to the minimum energy solution, which grows more rapidly than the gain achieved for the amount of layers explored. Without this varying scaling factor, the energy difference between both solutions increases with problem size.\\

\begin{figure}[!h]
\centering 
\includegraphics[width=\columnwidth]{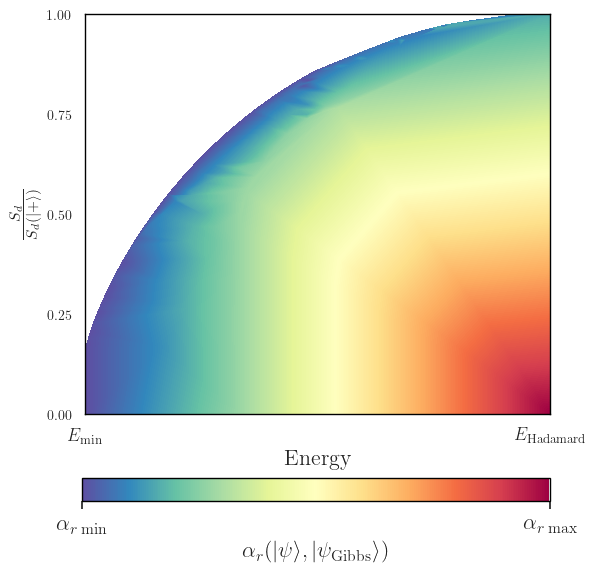} 
\caption{Heat map for a 6-node instance of the Max Cut problem. For each state located within the diagram, the associated color indicates the value of the relative quality of the indicated state always compared against the pure Gibbs state of equal energy $\alpha_{r}(\ket{\psi(E(\psi), S_{d}(\psi))}, \ket{\psi_{\text{Gibbs}}})$, i.e., the boundary state located on its same vertical.}
\label{fig:heat_map} 
\end{figure}

In Figure \ref{fig:states_vs_gibbs_states} (b) the comparison is performed for pure Gibbs states of different temperatures. In this case, initializing the QAOA with lower-temperature pure Gibbs states is shown to be the preferred strategy, with gains that appear to be supra-linear. Notably, and as explained in the methods section, reaching this lower temperature states via TNs means is also supra-linear in computational cost. However, the optimal Temperature at which we stop the TN algorithm depends on the capacity of both quantum and classical resources. Finally, Figure \ref{fig:states_vs_gibbs_states} (c) shows that as we approach the Boltzmann limit by creating higher entropy states at a fixed energy, the quality of the solution given by the QAOA algorithm improves. \\

The aforementioned results can be extended to a more in-depth comparison of possible states for a single instance of a random problem, as illustrated in Figure \ref{fig:heat_map}. To generate it, and following the depiction of the energy-entropy diagram, we sweep the energy range for a given instance, generating pure Gibbs states and random states at different levels of entropy and using them as initial states for the QAOA. Notably, the resulting coloring displays the relative difference between solutions obtained from states at the same energy verticals, using the pure Gibbs state initialized QAOA as reference energy, and is obtained through interpolation over the sampled results. Again, states close to or at the Boltzmann boundary are the best performing and results are consistent across different instances, albeit only a single one is shown for brevity.

\subsection{Relative coherence entropy analysis}

\begin{figure}[!t]
\centering 
\includegraphics[width=\columnwidth]{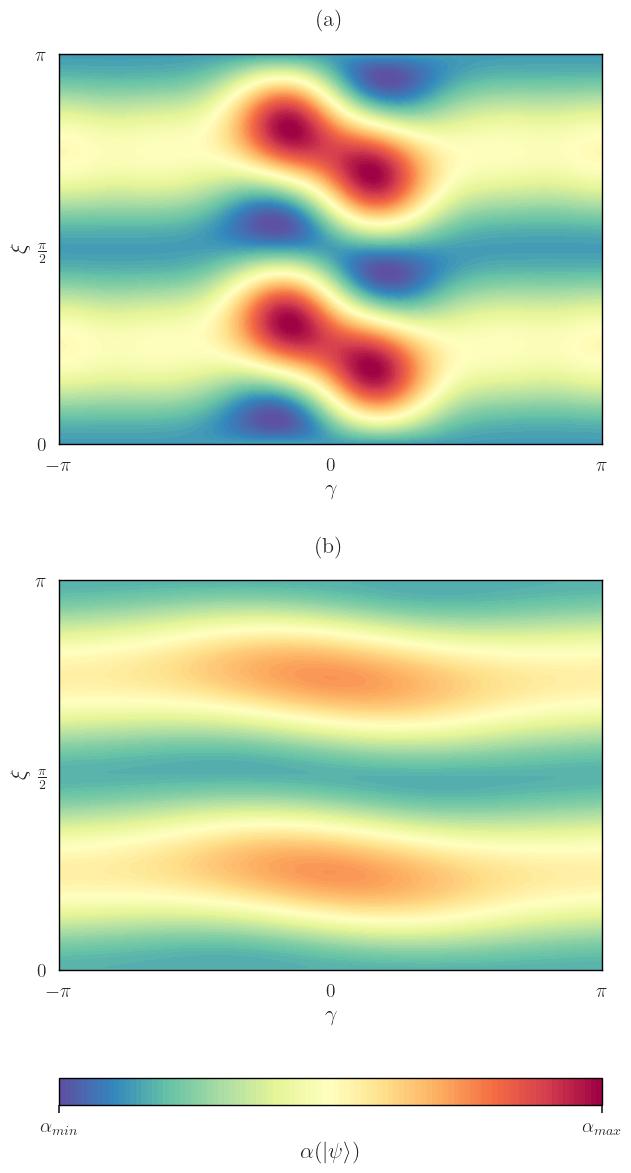} 
\caption{Landscape of energies generated by a $p=1$ QAOA layer for two different initialization states for an 8-node Max Cut problem. The angles $\gamma$ and $\xi$ correspond to the rotation parameters of the Cost and Mixer operators for a QAOA layer. The color within each plot represents the energy obtained by QAOA for those fixed parameters. (a) Landscape of energies generated with a pure Gibbs initialization state of temperature $T=3$. (b) Landscape of energies generated with a Gaussian state that has the same energy as the pure Gibbs state but with lower entropy.}
\label{fig:landscape_gibbs_vs_gauss} 
\end{figure}

Figure \ref{fig:landscape_gibbs_vs_gauss} shows the disparity in the landscape of energies that a single layer of the QAOA algorithm is able to generate for initialization states with almost equal energy (below 2$\%$ difference) but very different coherences (18\% difference). This suggests that, among the many properties of the optimization landscape generated, one of the advantages of initializing the QAOA algorithm with coherent quantums states is that it gains access to better minima and, hence, better solutions to which the optimizer can converge. In fact, the use of pure Gibbs states not only allows access to lower energy states, but also allows finding higher energy states.\\

Since this increased variability can not be linked to the commonly discussed expressiveness of the ansatz \cite{du_2022, funcke_2021}, as it is the same in both cases, we adapt this concept through the definition of a relevant measure  to quantify expressiveness for optimization tasks. In broad terms, this measure is linked to the effective volume of the Hilbert space that is explored by the ansatz for a given initial state, i.e., we aim to quantify the amount of states that generate significantly different probability distributions of solutions for the optimization task. To formalize this, we introduce the definition of the manifold that includes this relevant states.

\begin{definition}
[Manifold of $\epsilon$-close reachable states]
\label{def:expressivity-manifold}
Given a parametrized quantum circuit, $U(\vec \varphi)$, 
acting on an initial state $\ket{\psi_0}$, we define the manifold of $\epsilon$-close reachable states as the set of states that are $\epsilon$-close to a state that can be generated by the PQC. More formally,
\begin{equation}
\begin{aligned}
\mathcal{M}_\epsilon\left(U, \ket{\psi_0}\right) 
&=\{ \ket{\psi}\in \mathcal{H} \ : \ \exists \ \vec \varphi \in \mathbb{R}^n \ \\
&\quad | \ D\left(\ket{\psi}, U(\vec\varphi)\ket{\psi_0}\right)\leq \epsilon  \}
\end{aligned}
\end{equation}
where $D\left(\ket{\psi}, \ket{\phi}\right)$ is a distance between two states $\ket{\psi}$ and $\ket{\phi}$ of the Hilbert space.
\end{definition}
Under this definition, a quantum state maximally contributes to the volume of this manifold when the resulting probability distribution in the relevant basis is at least at distance $\epsilon$ from all the other states generated by the combination of the ansatz and the initial state. Our intuition is that initializing the QAOA with coherent states results in a larger volume of this manifold, which in turn would justify the appearance of deeper maxima or minima on average, by the fact that a larger volume of the space of all possible probability distributions would be approximately covered. \\

To validate our intuition, we resort to the use of Principal Component Analysis (PCA) \cite{shlens_2014, gewers_2021}. This technique allows us to reduce the dimensionality of a data set by projecting it into the most relevant or highest variability subspace. In more detail, we approximate the calculation of the volume of the relevant manifold for a given problem instance and initial state in the following manner:
\begin{itemize} 
\item  Sweep across all possible values of the parameters for the QAOA and obtain the resulting probability distribution.
\item Perform the PCA on the previous data set and keep the cross-section with the largest variance.
\item Compute the area of the envelope of such cross-section.
\end{itemize}

Hence, the calculation of the volume of the subspace is estimated by an area calculation (for the specific details on each of the steps, see Appendix \ref{app:pca}). \\

We employ this procedure through multiple randomized instances and initial quantum states with varying relative coherence entropy. For each initialization state and instance, we use the QAOA with a single layer ($p=1$) to generate the manifold of reachable states. The maximum dimension for this manifold is two, as a single layer of the QAOA only contains two parameters, $(\gamma, \xi)$. By applying the PCA, we project the 2-dimensional manifold, originally contained in a $k$-dimensional space ($k = 2^n$ for $n$-qubits), onto a 2-dimensional subspace of maximum variability. \\

\begin{figure}[!h]
\centering 
\includegraphics[width=\columnwidth]{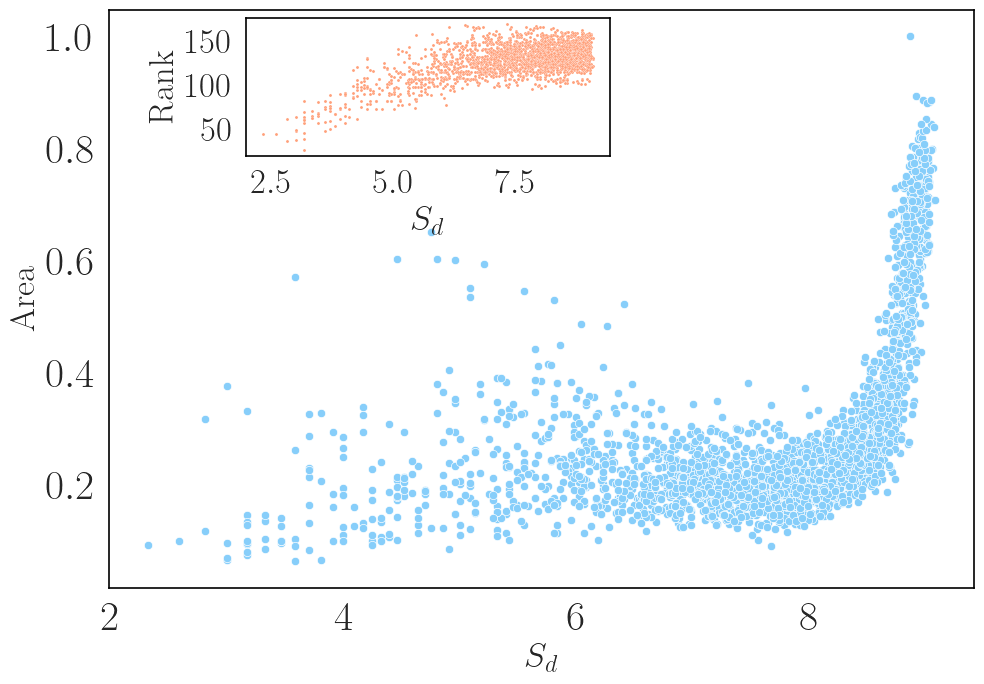} 
\caption{Comparison for 10-node Max Cut of the normalized area of the projection of the state space generated by a QAOA layer onto the two-dimensional subspace of maximum variability, as a function of the entropy of the initialization state. The area is normalized with respect to the instance that produces the largest area. The inset figure shows the Schmidt rank as a function of the entropy of the initialization state.}
\label{fig:entropy_area_rank} 
\end{figure}

Figure \ref{fig:entropy_area_rank} shows the results of the described experiments. We observe that, indeed, a higher coherence entropy of the initial state in the QAOA results in a larger estimated area of the relevant space generated by the algorithm. Additionally, as illustrated in the inset of Figure \ref{fig:entropy_area_rank}, a higher entropy in the initialization state correlates with a higher Schmidt rank, which indicates an increase in the number of principal directions with significant variability. Therefore, higher coherence entropy leads not only to a larger state subspace in terms of the area projected onto the directions of greatest variability, but also to an increased number of directions exhibiting large variability.

\subsection{Classical-quantum optimization}

\begin{figure}[!h]
\centering 
\includegraphics[width=1\columnwidth]{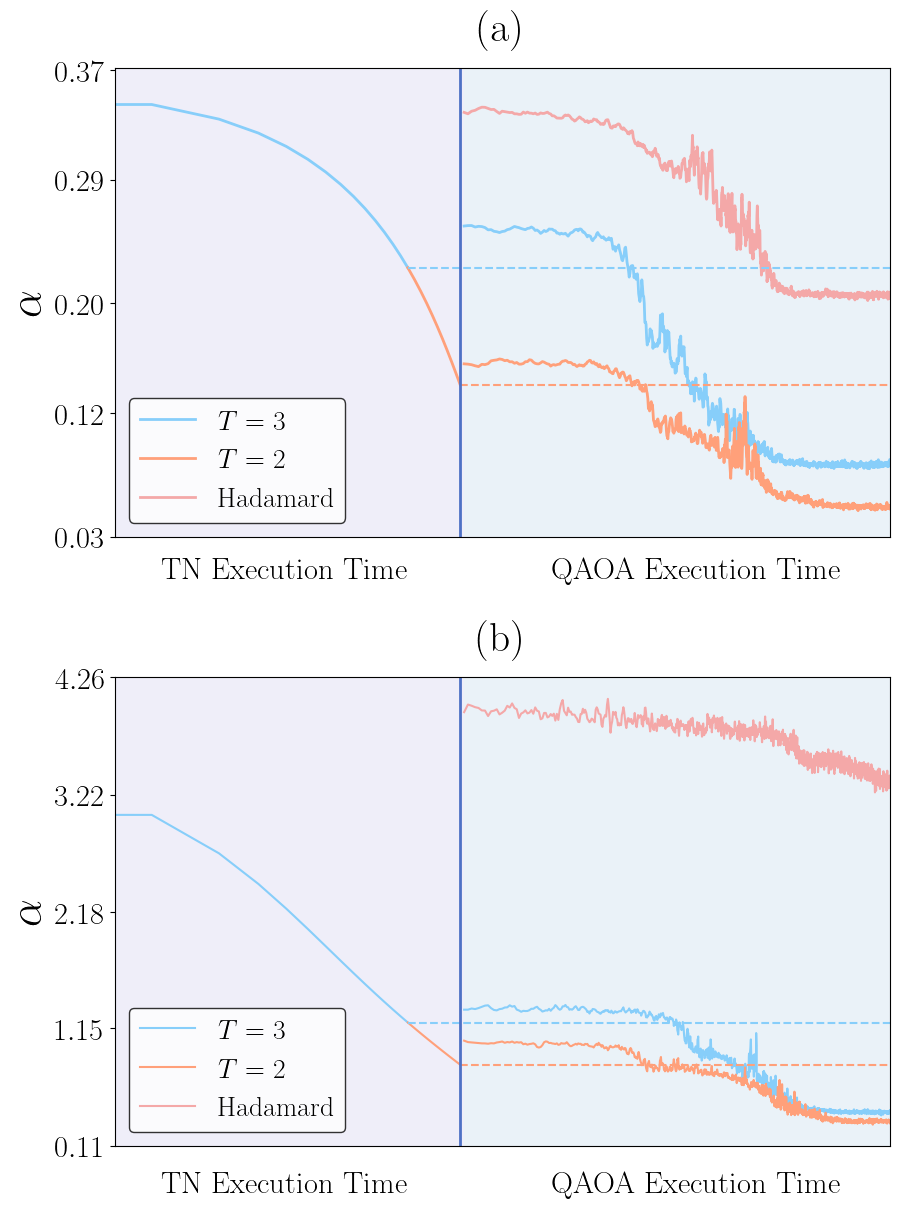} 
\caption{Quantum-classical optimization scheme for classical problems. In the first stage of the optimization (purple background), $\text{MPO} \; W^{\text{I,II}}$ constructs pure Gibbs states at different temperatures. In the second stage (blue background), the QAOA algorithm starts from the state generated and continues the optimization process. (a) 10-node random Max Cut problem (10 qubits). (b) 3-city random TSP problem (9 qubits).}
\label{fig:mpo_time_evolution_qaoa} 
\end{figure}

The combined optimization of the Tensor Networks $\text{MPO} \; W^{\text{I,II}}$ algorithm and the QAOA algorithm creates a classical-quantum optimization protocol, which reduces the required quantum computational resources while enhancing the performance of the variational quantum QAOA algorithm. We present results demonstrating the capacity of this quantum-classical scheme for random Max Cut and TSP \cite{ketg_2022} problems. Figure \ref{fig:mpo_time_evolution_qaoa} illustrates the typical behavior of the classical-quantum optimization process. In the initial stage, an imaginary time evolution of the system is performed, allowing the construction of pure Gibbs states at a specific temperature. Subsequently, the QAOA algorithm utilizes the state generated by the $\text{MPO} \; W^{\text{I,II}}$ as its initialization quantum state. \\

Notably, the state obtained through $\text{MPO} \; W^{\text{I,II}}$ and the initial state of the QAOA algorithm exhibit a slight difference in terms of energy. This discrepancy arises because the conversion process from the pure Gibbs state, represented as an MPS, to a PQC is not exact. The small difference in fidelity between the starting MPS state and the resulting PQC state results in a minor shift in the starting point with respect to energy. The results discussed in section \ref{subsec:results_gibbs_states} can be visualized in this context. Initializing the QAOA algorithm with a lower-temperature pure Gibbs state enables it to reach higher-quality quantum states at fixed depth. Moreover, the initial optimization phase helps to prevent the QAOA algorithm from being trapped in higher energy local minima. \\

Our preliminary tests indicate that our approach is also advantageous when using different types of ansatz (in particular, the SU(4)-based ansatz suggested in \cite{rudolph_2023}) and when applied to larger scale instances of more complex and realistic optimization problems. However, a more detailed and extensive analysis is required. The presented results demonstrate the potential of the designed quantum-classical scheme for optimization tasks, even when focusing on limiting the amount of quantum resources used to maximize its short-term utility.

%% file: sec/05-acknowledgements.tex
\section{Acknowledgments}
This work was supported by the Spanish
CDTI through Misiones Ciencia e Innovación Program (CUCO) under Grant MIG20211005. 
The authors would like to thank the Theory team at Qilimanjaro, Mikel Garcia de Andoin, Julian Ferreiro, and the rest of the Tecnalia team for helpful discussions. J.R thanks A.G-S and Professor R.Rey for their very helpful advice.

%% file: sec/06-data_availability.tex
\section{Data availability}

All necessary data as well as the code implemented to evaluate the results and conclusions of this work are available from the corresponding author upon reasonable request.

%% file: sec/07-author_contributions.tex
\section{author contributions}

A.N.C, J.R, and A.R conceived and developed the main theoretical framework, designed the experiments and analysed the main results presented in the article. A.N.C performed the numerical simulations. J.B implemented the tensor networks code of the MPO Time Evolution algorithm. A.N.C, J.R, and J.M implemented the MPS to PQC translation code. A.N.C, J.R, and J.M worked on the analytical version of the algorithm, and A.N.C on the iterative version. P.T implemented the DMRG algorithm of the synergistic training framework and assisted in the initial proofs of concept. All authors contributed to the final version of the manuscript.

%% file: main.bbl
\providecommand{\noopsort}[1]{}\providecommand{\singleletter}[1]{#1}%

%% file: sec/08-appendix.tex
\appendix

\section{Synergistic training for classical problems}
\label{app:synergetic_training}

For the synergistic optimization scheme presented in Ref. \cite{rudolph_2023} the criterion for selecting favorable initialization states is given by the minimization of the energy expression

\begin{equation}
\label{eq:energy_exprected_value}
        E(\ket{\psi}) = \bra{\psi}H\ket{\psi}.
\end{equation}

Using the state energy criterion to search for candidate states for the case of classical problems, the TN algorithm employed in the initial optimization, the Density Matrix Renormalization Group (DMRG) \cite{schollwck_2005} tends to produce candidate states for initialization that are typically close to the states in the computational basis. This is because for classical problems such as the Max Cut problem \cite{lalovic_2024} or Traveling Salesman Problem (TSP) \cite{ketg_2022} problem, the solutions usually correspond to eigenstates of the Hamiltonian operator that are directly represented by product states in the computational basis. This brings a major problem because a VQA initialized with a state that is a state in or near the computational basis has limited potential for further improvement, thus reducing the efficiency of the optimization process. \\

\begin{figure}[!h]
\centering 
\includegraphics[width=\columnwidth]{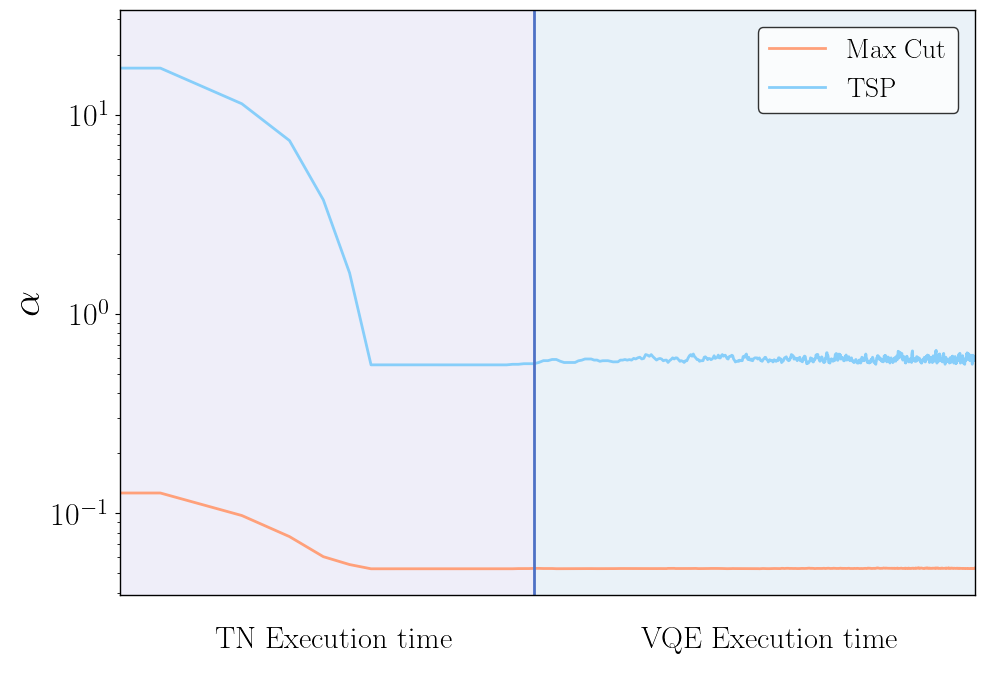} 
\caption{Optimization process using the protocol of Ref. \cite{rudolph_2023} for classical problems. First, the Density Matrix Renormalization Group (DMRG) algorithm is used to optimize the state for classical problems. In the second step, a Variational Quantum Eigensolver (VQE) optimization is performed using the state obtained from DMRG as a starting point, with a linear ansatz structure. For both problems, the Max Cut problem with 14 nodes (14 qubits) and the TSP with 3 cities (9 qubits), the DMRG algorithm converges to states close to the computational basis states. If the starting state generated by the DMRG algorithm is not used, the VQE algorithm can find states of lower energy.}
\label{fig:dmrg_vqe_training} 
\end{figure}

This behavior has been consistently observed when the synergistic optimization protocol has been used to solve classical optimization problems. Figure \ref{fig:dmrg_vqe_training} shows the behavior described in the previous paragraph for two randomized problems, each belonging to different classes of classical problems. The behavior shown in Figure \ref{fig:dmrg_vqe_training} is associated with the fact that the eigenstates of the Hamiltonian correspond directly to states of the computational basis. The DMRG algorithm performs an iterative diagonalization process by which it attempts to find a state of the Hamiltonian that minimizes the equation \ref{eq:energy_exprected_value}. This process usually leads irremediably to states of the computational basis or to states close to those of the computational basis.

\section{Implementation of $\text{MPO} \; W^{\text{I,II}}$}
\label{app:mpo_time_evolution}

The goal is to obtain an MPO that is able to reproduce the time-evolving operator presented in (\ref{eq:taylor_evolution_operator}). However, since there isn't any efficient way to construct an MPO representing the time-evolving operator, the two approximations presented in \cite{zaletel_2015} have been implemented.

\subsection{Non-overlapping approximation}

One of the simplest methods to consider is to take the Euler time-stepper $\prod_{x}(1 + \delta \tau H_{x})$. However, in order to have an exact MPO representation, performing a modified Taylor expansion is required such that we end with the approximate evolution operator of the form

\begin{equation}
\begin{aligned}
U^I(\delta\tau) &= 1 + \delta\tau \sum_{x} H_x  + \delta\tau^2 \sum_{x<y} H_x H_y \\
&\quad + \delta\tau^3 \sum_{x<y<z} H_x H_y H_z + \ldots,
\end{aligned}
\label{eq:first_taylor_evolution_operator}
\end{equation}

where $x < y$ denotes the terms of $H_x$ that lie strictly to the left of the site affected by $H_y$. 
Therefore, it can be seen that terms only include operators $H_i$ that do not overlap between each other in any site. Expression \ref{eq:first_taylor_evolution_operator} has a straightforward MPO representation, where for each site $i$ the corresponding tensor is given by

\begin{equation}
W^{I}_{i}(\delta\tau) =
\begin{blockarray}{rcc}
& \matindex{1} & \matindex{$N_{i}$} \\
\begin{block}{r(cc)}
\matindex{1 \;} & {\mathds{1}}_{i} + \delta\tau {D}_{i} & \sqrt{\delta\tau}{C}_{i} \\
\matindex{$N_{i-1}$ \;} & \sqrt{\delta\tau}{B}_{i} & {A}_{i} \\
\end{block}
\end{blockarray}
\; \; \; .
\label{eq:mpo_first_order}
\end{equation}

Elements $A_i, B_i, C_i, D_i$ can be obtained according to Eq. (\ref{eq:mpo_hamiltonian}). The error introduced in the approximation $U(\delta\tau)\rightarrow U^I(\delta\tau)$ scales as $\mathcal{O}(L\delta\tau^2)$.

\subsection{Single-site overlapping approximation}

It is possible to get a more accurate approximation of (\ref{eq:taylor_evolution_operator}) by allowing the subsets in which the local operators $H_i$ act to overlap by at most one site in each term. This operator will take the form

\begin{equation}
\begin{aligned}
U^{II}(\delta\tau) &= 1 + \delta\tau \sum_{x} H_x  + \frac{\delta\tau^2}{2} \sum_{x,y} H_x H_y \ldots , 
\end{aligned}
\label{eq:second_taylor_evolution_operator}
\end{equation}

where in each term, the intersection between the subsets $x, y$ cannot be larger than one site. \\

Strictly the error is still $\mathcal{O}(L\delta\tau^2)$, but as shown in \cite{zaletel_2015} it will typically be much lower than the one made in (\ref{eq:first_taylor_evolution_operator}). In this case building the MPO representation $W^{II}$ will bring an additional error of $\mathcal{O}(\delta\tau^3)$, however it will not be very relevant compared with the error made in the approximation (\ref{eq:second_taylor_evolution_operator}).\\

$W^{II}$ can be divided into sub-blocks such that for each site $i$ we have

\begin{equation}
W^{II}_{i} =
\begin{blockarray}{rcc}
& \matindex{1} & \matindex{$N_{i}$} \\
\begin{block}{r(cc)}
\matindex{1 \;} & W^{II}_{D_i} & W^{II}_{C_i} \\[0.1cm]
\matindex{$N_{i-1}$ \;} & W^{II}_{B_i} & W^{II}_{A_i} \\
\end{block}
\end{blockarray}  \; \; \; ,
\label{eq:mpo_second_order}
\end{equation}

where the tensor elements $W^{II}_{S_i;j,k}$ with \mbox{$S_i\in \{A_i,B_i,C_i,D_i\}$} can be extracted using the expression

\begin{widetext}
\begin{equation}
W_{S_i ; j,k}^{II} = \left(\begin{array}{llll}
\delta_{S_i, D_i} & \delta_{S_i, C_i} & \delta_{S_i, B_i} & \delta_{S_i, A_i}
\end{array}\right) \exp \left\{\left(\begin{array}{cccc}
\delta\tau D_i & 0 & 0 & 0 \\
\sqrt{\delta\tau} C_{i ; 1, k} & \delta\tau D_i & 0 & 0 \\
\sqrt{\delta\tau} B_{i ; j, 1} & 0 & \delta\tau D_i & 0 \\
A_{i ;j,k} & \sqrt{\delta\tau} B_{i ; j,1} & \sqrt{\delta\tau} C_{i ; 1,k} & \delta\tau D_i
\end{array}\right)\right\}\left(\begin{array}{l}
1 \\
0 \\
0 \\
0
\end{array}\right)
.
\label{eq:mpo_second_order_elements}
\end{equation}
\end{widetext}

\section{$\text{MPO} \; W^{\text{I,II}}$ for Gibbs States}
\label{app:mpo_gibbs_states}

The quality of the solution obtained by $\text{MPO} \; W^{\text{I,II}}$ for constructing Gibbs states depends on several factors: the total evolution time $t$, the discretization $\delta \tau$ used, and the internal dimension $\chi_{max}$ allowed for both the MPS representing the quantum state and the MPO representing the Hamiltonian. This bond dimension is the one used to truncate after each application of an MPO layer, such that the dimension of the MPS remains constant during the entire evolution. The truncation method that minimizes the introduced error, and which has been used in this work, is known as canonical truncation \cite{cuevas_2017}. \\

The imaginary time evolution over the state $\ket{+}^{\otimes n}$ generates a distribution such that for each eigenstate of the Hamiltonian, its corresponding amplitude is given by

\begin{equation}
c_{s} = \frac{e^{- \frac{E_{s}}{T}}}{\sqrt{Z}},
\label{eq:coeficient_gibbs}
\end{equation}

where the factor $Z$ represents the partition function of the system and serves to normalize the state of the system, $T$ denotes the associated temperature, and $E_{s}$ is the energy associated to the eigenstate $\ket{s}$ of the Hamiltonian. Taking the natural logarithm, the equation \ref{eq:coeficient_gibbs} becomes the expression 

\begin{equation}
\ln(c_{s}) = \frac{- E_{s}}{T} - \ln(\sqrt{Z}).
\label{eq:coeficient_gibbs_log}
\end{equation}

\begin{figure}[!h]
\centering 
\includegraphics[width=\columnwidth]{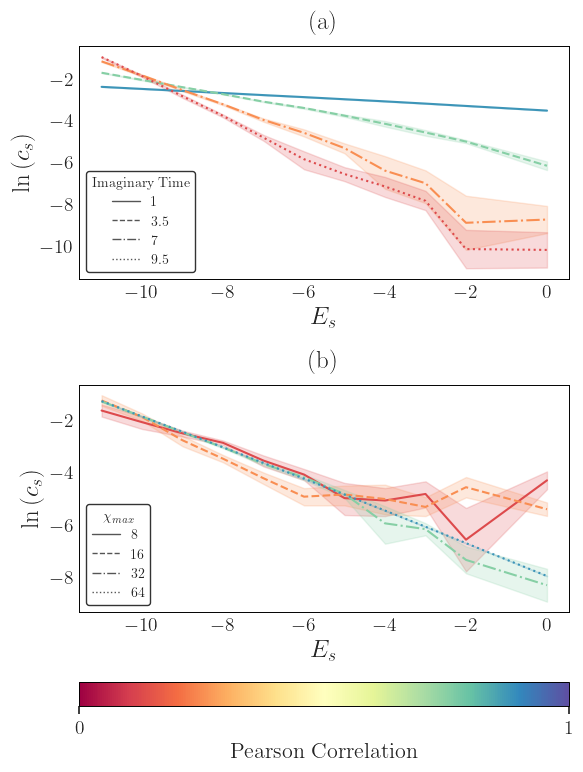} 
\caption{Amplitudes distributions illustrating the quality of the Gibbs states generated by $\text{MPO} \; W^{\text{II}}$ for an 8-qubit system. The color coding in both graphs indicates the quality of the solutions generated using Pearson's correlation: red represents the lowest quality and blue represents the highest quality. (a) For the same Hamiltonian, imaginary time evolution is performed for different times using a constant time step $\delta \tau = 0.01$ and a fixed internal dimension $\chi_{max}=32$. (b) For the same Hamiltonian and with $t=6$ and $\delta \tau = 0.01$ fixed, imaginary time evolution is carried out with varying values of $\chi_{max}$.}
\label{fig:gibbs_states_from_mpo} 
\end{figure}

The linear behaviour shown in expression \ref{eq:coeficient_gibbs_log} can be used to check the quality of the numerically generated Gibbs states. After sampling and plotting them as shown in Figure \ref{fig:gibbs_states_from_mpo} we expect a straight line with a slope of $-\frac{1}{T}$. In order to extract samples from an MPS we can use algorithms such as the one presented in \cite{ferris_2012}.\\

\section{Principal Component Analysis for area estimation}
\label{app:pca}

\begin{figure*}[!th]
\centering 
\includegraphics[width=2\columnwidth]{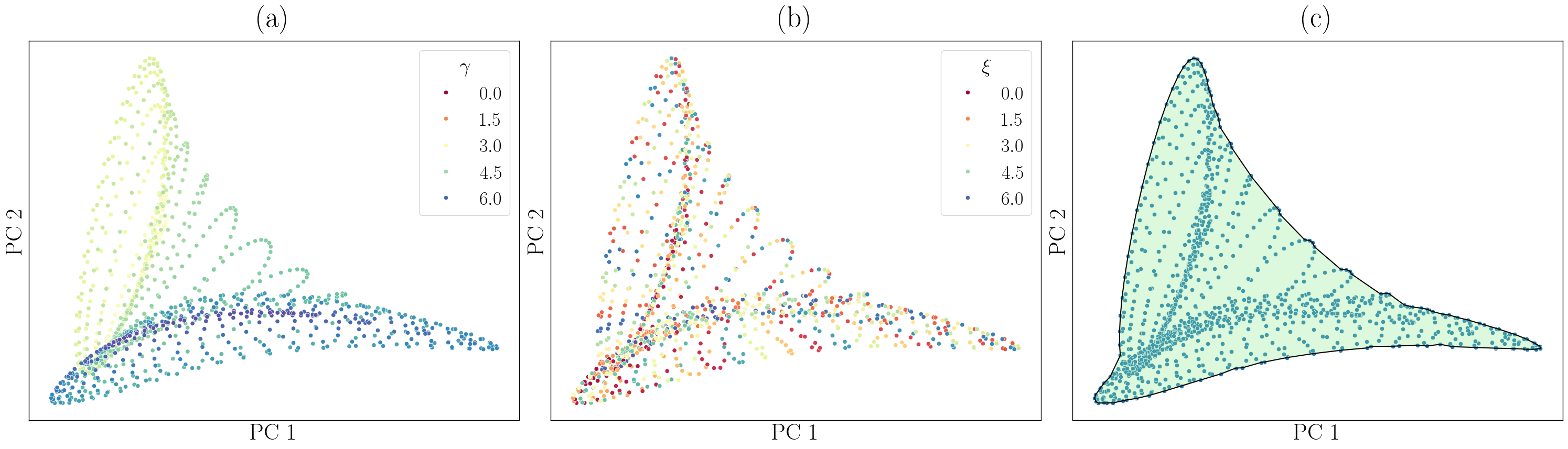} 
\caption{Projection onto the 2-dimensional subspace defined by the principal vectors of maximum variability for the states generated by a $p=1$ QAOA layer, with the initialization state being a Gibbs state at temperature $T=5$, for a 6-node random Max Cut problem. (a) States generated in the 2-dimensional projection as a function of the $\gamma$ parameter of the cost operator in the QAOA. (b) States generated in the 2-dimensional projection as a function of the $\xi$ parameter of the mixer operator in the QAOA. (c) Calculation of the area using the concave envelope of the set of projected states in the 2-dimensional subspace.}
\label{fig:states_subspace} 
\end{figure*}

For a given depth of the QAOA, one can obtain a sample of the manifold of states generated from a specific initialization state by sweeping across the values that the parameters can take (in an equispaced manner). The larger the degree of granularity, the more precise the estimation of the volume  of the manifold, but also the more expensive computationally. From these states, a matrix $M$ of data can be constructed, such that $M_{ij}$ corresponds to the probability of measuring the element $j$ of the computational basis for a given state $(\vec{\gamma_i}, \vec{\xi_i})$ \\

However, even for the simplest QAOA ansatz possible, consisting of a sigle layer, it is extremely challenging to compute its area as it is embedded in a space of very large dimension. To address this complexity, we apply Principal Component Analysis to reduce the problem's dimensionality from an $n$-dimensional space to a 2-dimensional one. \\

The PCA is a well-known dimensionality reduction technique which is based in identifying the principal components of a data distribution, i.e., the basis for which the covariance matrix of the distribution is diagonal. From this basis, one can identify the subspace of Principal Components that captures the largest amount of variance of the data and reduce the dimensionality of the dataset by projecting it into such subspace. \\

This technique allows us to transform the volume calculation problem into a more manageable area calculation problem, albeit by the implicit assumption that the generated cross-section is representative of the relevant volume we want to compute.  This is done by applying the PCA to matrix $M$ and projecting the data into the subspace of the first two Principal Components (assuming that they are ordered decreasingly by variance captured). Figure \ref{fig:states_subspace} (a) and (b) illustrates the distribution of points associated with the projections of the states onto the subspace defined by these two principal vectors of maximum variability for a given random instance. Since the state space is generated by a QAOA layer that depends on two parameters, $\gamma$ and $\xi$, the projection effectively reduces a 2-dimensional manifold embedded in an $n$-dimensional space to a 2-dimensional subset within a 2-dimensional space. In addition, as shown in Figure \ref{fig:states_subspace} (a) and (b), one can observe the impact of each of the parameters in the generated trajectories, with each ring corresponding to a specific value of $\gamma$ and being traversed by varying the values of $\xi$. Once the projection onto the 2-dimensional subspace is completed, we can proceed, as illustrated in Figure \ref{fig:states_subspace} (c), with calculating the area associated with the projected points. This area calculation is performed using the concave envelope method \cite{chadnov_2004}, where we seek the optimal envelope that encloses the set of points in order to estimate the interior area. The calculation of the concave envelope was carried out using the \textit{Alphashape} package.